\documentclass[twocolumn,showpacs,preprintnumbers,superscriptaddress,amsmath,amssymb]{revtex4}

\usepackage{epsfig}
\usepackage{graphicx}
\usepackage{dcolumn}
\usepackage{bm}
\usepackage{times}
\usepackage{xcolor}

\begin{document}

\title{Suppressed epidemics in multi-relational networks}

\author{Elvis H. W. Xu}
\affiliation{Department of Physics, The Chinese University of Hong
Kong,\\ Shatin, New Territories, Hong Kong SAR, China}

\author{Wei Wang}
\affiliation{Web Sciences Center, University of Electronic Science
and Technology of China, Chengdu 610054, China}

\author{C. Xu}
\affiliation{College of Physics, Optoelectronics and Energy,
Soochow University, Suzhou 215006, China}

\author{Ming Tang}
\affiliation{Web Sciences Center, University of Electronic
Science and Technology of China, Chengdu 610054, China}

\author{Younghae Do}
\affiliation{Department of Mathematics, Kyungpook National
University, Daegu 702-701, South Korea}

\author{P. M. Hui}
\affiliation{Department of Physics, The Chinese University of Hong
Kong,\\ Shatin, New Territories, Hong Kong SAR, China}
\date{\today}

\begin{abstract}
A two-state epidemic model in networks with links mimicking two
kinds of relationships between connected nodes is introduced.
Links of weights $w_{1}$ and $w_{0}$ occur with probabilities $p$
and $1-p$, respectively.  The fraction of infected nodes $\rho(p)$
shows a non-monotonic behavior, with $\rho$ drops with $p$ for
small $p$ and increases for large $p$.  For small to moderate
$w_{1}/w_{0}$ ratios, $\rho(p)$ exhibits a minimum that signifies
an optimal suppression. For large $w_{1}/w_{0}$ ratios, the
suppression leads to an absorbing phase consisting only of healthy
nodes within a range $p_{L} \le p \le p_{R}$, and an active phase
with mixed infected and healthy nodes for $p < p_{L}$ and
$p>p_{R}$.  A mean field theory that ignores spatial correlation
is shown to give qualitative agreement and capture all the key
features.  A physical picture that emphasizes the intricate
interplay between infections via $w_{0}$ links and within clusters
formed by nodes carrying the $w_{1}$ links is presented. The
absorbing state at large $w_{1}/w_{0}$ ratios results when the
clusters are big enough to disrupt the spread via $w_{0}$ links
and yet small enough to avoid an epidemic within the clusters.  A
theory that uses the possible local environments of a node as
variables is formulated. The theory gives results in good
agreement with simulation results, thereby showing the necessity
of including longer spatial correlations.

\end{abstract}

\maketitle

\section{Introduction}
The past 15 years have witnessed the rapid development in complex
network science and its
applications~\cite{Albert2002,Dorogovtesv2008}.  Besides
structural properties, how network topology, dynamical behavior
and functionality are coupled together has been a focus of
research and much work has been done within a single complex
network~\cite{Dorogovtesv2008,Pastor-Satorras2014,Castellano2009}.
Recent works have revealed that the heterogeneous nature of the
nodes and links, often interpreted as multilayer networks,
inter-dependent networks and interconnected networks, is crucial
for understanding the properties of real-world complex systems
such as social networks, complex infrastructures, and brain
networks~
\cite{Robert2008,Bagheri2009,Bullmore2009,Kivela2014,Gao2012,Boccaletti2014}.
With a network with a specific function, e.g. for information
flow, power grid, vehicular traffic, and air traffic, representing
a layer, the proper functioning of an entire complex system is
related to the simultaneous operation of an interacting set of
networks. Recent studies alone this line include empirical
analysis of real-world network data~\cite{Szell2010,
Cardillo2013}, evolution of network
structures~\cite{Battiston2014,Nicosia2013,Kim2013}, and new
critical phenomena and processes occurring on
them~\cite{Gao2012,Boccaletti2014,Buldyrev2010,Granell2013}. These
coupled networks exhibit some common features, such as the inter
degree-degree correlation~\cite{Lee2012},
inter-similarity~\cite{Parshani2010}, multiple dependence in
providing support~\cite{Gao2012}, and node and edge overlapping
between layers~\cite{Cellai2013}.  These features have important
effects on critical phenomena and the dynamics, including
percolation~\cite{Gao2012}, cascading failure~\cite{Buldyrev2010},
diffusion processes~\cite{Gomez2013}, emergence of
cooperation~\cite{Gomez-Gardenes2012} and epidemic
dynamics~\cite{Granell2013}, when compared with those in a single
network.  For example, Buldyrev \emph{et al.} found that strong
structural heterogeneity increases the vulnerability of multilayer
networks to random failure, an effect opposite to that in a single
network~\cite{Buldyrev2010}.  For spreading dynamics in layered
networks, Wang and coworkers found that the degree-degree
correlation between the layered networks cannot change the
information threshold, but make the system more resilient to
epidemic outbreak~\cite{Wang2014}.

In the present work, we focus on multi-relational networks.  The
simplest form of it refers to a set of nodes that are connected by
two or more types of links, signifying links of different
importance or relationships.  They can also be regarded as a type
of multilayer networks, with the same set of nodes connected
differently in each layer representing, for example, different
social relationships~\cite{Kivela2014}.  The relationships could
be friends, relatives, colleagues, clients, and schoolmates, among
others, in the context of a social network~\cite{Cai2005}; or
enemies, guild members, and friends in online
games~\cite{Szell2010}.  The same cities on a map could be
connected by different links representing highways, railways and
airline routes~\cite{Gu2011}.  The different links undoubtedly
play different roles in spreading dynamics.  Our everyday
experience is that we would be selective in sharing our views or
our latest updates, and would not share them equally with everyone
we know.  Often, individuals are more willing to share good news
with their friends~\cite{Yang2008}.  Two friends could communicate
via several ways.  The means of communication often reflects the
urgency of the contents and phone calls are preferred over short
messages in conveying urgent information to
friends~\cite{Holme2012}.

In epidemics, the chance of infecting another person is not even
among all the ones that an infected person comes into contact.
Instead, an infected person will have a higher chance of infecting
another person who is in closer and longer contact, e.g. a
colleague in the same office.  The different importance among the
links in a network with multi-relations gives rise to an uneven or
biased chance of someone being infected.  Here, we aim to study
the effects of this biased selection on the spreading dynamics.  We
propose and study a model that captures the non-trivial effects of
two different kinds of links in a random regular network of degree
$k$. A fraction $p$ of the links carry a higher
weight $w_{1}$ and the remaining $1-p$ carry a weight $w_{0}$. A link
of higher weight has a higher chance to be used as a path for
infection~\cite{Yang2012,Wang2014a,Min2013}.  Within the
susceptible-infected-susceptible (SIS) model of
epidemics~\cite{Castellano2010,Boguna2013,Lee2013,Cohen2010},
we study the extent
of an epidemic as measured by the fraction of infected sites
$\rho$ in the steady state as a function of $p$ and the
contrast $w_{1}/w_{0}$ in the weights in detail. It is found that
$\rho(p)$ exhibits a non-monotonic dependence, with $\rho$ drops with $p$
for small $p$ and increases for large $p$.
For small to moderate $w_{1}/w_{0}$ ratios, $\rho$ exhibits a minimum at
some value of $p$.
For large $w_{1}/w_{0}$ ratios, there exists a range of $p$ in
which the system carries 100\% healthy nodes. This leads to
a re-entrance behavior in which the system starts with an active and
mixed phase consisting of both infected and healthy nodes for $0 \le p
\le p_{L}$, makes a transition into an absorbing and all-healthy phase at
$p=p_{L}$ and stays as such within a range $p_{L} \le p \le p_{R}$, and
re-enters into an active and mixed phase for $p_{R}<p<1$.
To understand the behavior, we
report results of two analytic approaches together with a physical
picture.  We set up a single-site mean field approach that
captures all these features.  Despite the theory only gives
qualitative agreement with simulation results, it has the merit
that analytic expressions for $\rho(p)$ in the $p \rightarrow 0$
and $p \rightarrow 1$ limits can be derived.  It also gives
phase diagram that exhibits the re-entrance behavior when the
contrast $w_{1}/w_{0}$ is above a threshold.  A physical picture
that emphasizes the importance of the clusters formed by nodes
that carry links with the higher weight then emerges.  These links
serve to confine the spread.  At small $p$,
the cluster sizes are small and disease in such small clusters
cannot sustain.  Thereby, they serve as sinks for the disease and
effectively reduce the infection probability and lead to a drop in
$\rho$.  At large $p$, there is a big cluster and the disease can
be sustained within the cluster. Replacing some $w_{1}$ links by
$w_{0}$ links reduces effectively the infection probability and
$\rho$ drops from the $p=1$ limit. For sufficiently large
$w_{1}/w_{0}$, 100\% healthy phase is achieved for a range
of $p$ when the cluster sizes are big enough to disrupt the spread via
$w_{0}$ links and yet not so big for the disease to sustain
through infections in the clusters.  We further
constructed a more accurate theory that uses the different local
environments of a node as the variables~\cite{Noel}.  The theory,
with its longer spatial correlation, is shown to give results in
quantitative agreement with simulation results.

The plan of the paper is as follows. In Sec.~II, the model of
spreading dynamics in a network with links corresponding to two
kinds of relationship is defined.  Key features of $\rho(p)$ on
other parameters of the model as observed in detailed numerical
simulations are described in Sec.~III.  In Sec.~IV, we develop a
mean-field theory and show that the theory exhibits all the
observed features, although the agreement is only quantitative.  A
phase diagram that exhibits the re-entrance behavior is
constructed.  A physical picture on the role of the inhomogeneity
among the links is presented.  In Sec.~V, we describe the
construction of an improved theory based on considering the
dynamics of the local environment of a node.  The theory is shown
to give good agreement with simulation results.  We summarize the
work in Sec.~VI.

\section{Model}
Consider a network consisting of $N$ nodes.  A node $i$ is
connected to $k_{i}$ other nodes.  Each link that connects two
nodes $i$ and $j$ in the network carries a weight $w_{ij}$, which
is assigned when the network is constructed and its value remains
unchanged.  The different weights among the links represent
different kinds of relationship.  Within the context of an
epidemic model, each node, representing a person, can be in one of
two states: susceptible (S) or infected (I).  We study the effects
of inhomogeneous weighting of the links on a SIS epidemic model,
based on a contract
process~\cite{Varghese2013,Castellano2006,Munoz2010}. In a time
step, every infected node $i$ selects one of the $k_{i}$
neighbors, say node $j$, for a possible infection with the
probability
\begin{equation}
{\cal P}_{ij} = \frac{w_{ij}}{\sum_{j \in \{i\}} w_{ij}} \;,
\end{equation}
where the summation in the denominator is over the set $\{ i \}$
of $k_{i}$ nodes that are connected to the node $i$.  The
selection is thus biased by the weights of the links.  If the
selected neighbor is in the susceptible state $S$, then it will be
infected with an infection probability $\lambda$. Once infected,
its state becomes $I$.  If the selected neighbor is already in the
infected state $I$, then it will remain in state $I$.  A recovery
process is then carried out in which every infected node at the
beginning of the time step would recover to become state $S$ with
a recovery probability $\gamma$. The epidemic dynamics is then
repeated.  A quantity of interest is the fraction of infected
nodes in the system in the steady state. Note that for a
realization of the network, ${\cal P}_{ij}$ is a property of a
link and it does not evolve in time.

To study how the coexistence of links of different weights affects
the extent of an epidemic, we consider a distribution ${\cal
D}(w_{ij})$ in the weights among the links of the form
\begin{equation}
{\cal D}(w_{ij}) = (1-p)\, \delta(w_{ij} - w_{0}) + p\,
\delta(w_{ij} - w_{1}) \;.
\end{equation}
It represents a system with two types of relationship.  The
network has a fraction $1-p$ of the links carrying a weight of
$w_{0}$ and a fraction $p$ carrying a weight of $w_{1}$. A few
points should be noted from the expression of ${\cal P}_{ij}$ for
a realization of the weighted network.  The weights $w_{0}$ and
$w_{1}$ are used in preferentially selecting a neighbor and it is
the ratio $w = w_{1}/w_{0}$ that matters.  The cases in which a
network consisting only of a single type of links, i.e., $w_{0}$
links ($p=0$), $w_{1}$ links ($p=1$), and $w_{0}=w_{1}$ (all
values of $p$), are equivalent.

In numerical simulations, there are different sources of
randomness.  Even for a given degree distribution for the nodes,
the connections among the $N$ nodes vary in different
realizations.  In addition, the assignment of $w_{0}$ and $w_{1}$
to the links can also be different between realizations for a
given value of $p$, even if the links are fixed in a network. In
what follows, results from simulations are obtained by averaging
over 100 different realizations of network construction and weight
assignment.  The initial condition is that of a whole lattice of
infected nodes and results are recorded after the system has
evolved beyond the transient period.  Without loss of generality,
we set $w_{0}=1$ and vary $w_{1}$ for $w_{1}>1$. Thus $w > 1$.
The averaged fraction of infected nodes is denoted by $\rho$,
which is a function of $p$, $\lambda$, and $\gamma$.  In defining
the problem, we intentionally allow for the possibility of
studying different types of networks as given by a distribution in
the degrees $k_{i}$ and different types of weight distribution
${\cal D}(w_{ij})$.

\section{Key Features in simulation results}

\begin{figure}
\begin{center}
\epsfig{figure=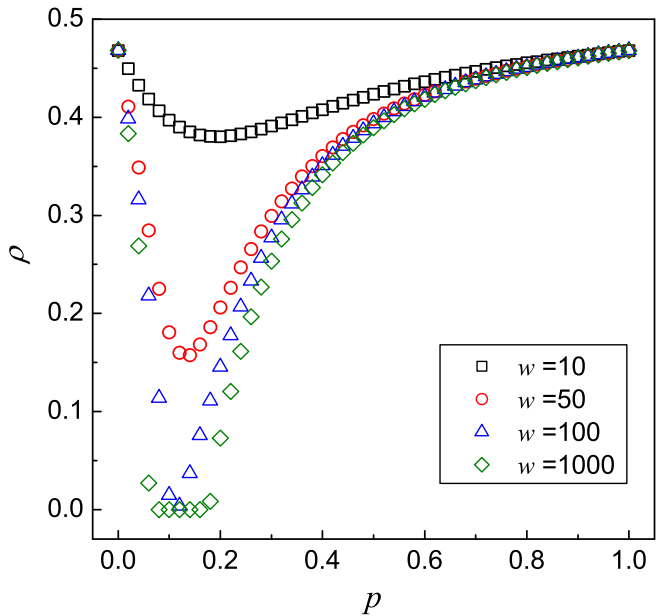,width=1.0\linewidth} \caption{The fraction
of infected nodes $\rho$ as a function of the fraction $p$ of
heavier weighted links.  Results are obtained by numerical
simulations in random regular networks of degree $k=10$. The SIS
model parameters are $\lambda = 0.02$ and $\gamma=0.01$. Results
for links with a weight ratio $w = w_{1}/w_{0} = 10$, $50$, $100$
and $1000$ are shown by different symbols with $w_{0}=1$.}
    \end{center}
\end{figure}
For concreteness, we focus on implementing the model in random
regular networks with nodes all having the same degree $k$.
Figure~1 shows the dependence of $\rho$ on the fraction of
weighted links $p$ for different values of $w=w_{1}/w_{0}$ in
networks with $N=5000$ nodes and $k=10$.  The SIS model parameters
are $\lambda = 0.02$ and $\gamma=0.01$. The equivalence of the
$p=0$ and $p=1$ cases necessarily leads to a non-monotonic
behavior of $\rho(p)$.  For a ratio of $w = 10$, the results are
typical with the key features that (i) a dilute fraction of
heavier weighted links suppresses $\rho$, (ii) $\rho(p)$ increases
with $p$ for a wide range of larger values of $p$, and (iii) there
is a value of $p$ at which $\rho$ is a minimum.  These general
features remain for higher $w_{1}/w_{0}$ contrast, as shown in the
results for $w_{1}/w_{0} = 50$ and $100$ in Fig.~1.

For $w_{1}/w_{0} \gg 1$, e.g. $w_{1}/w_{0} = 1,000$ (see Fig.~1),
a re-entrance behavior is observed.  The drop in $\rho(p)$ for
small $p$ is so strong that $\rho$ vanishes at a value of
$p=p_{L}$.  For a range $p_{L} \le p \le p_{R}$, $\rho = 0$
indicating the system evolves to a state with all nodes recovered
and remain healthy.  This corresponds to an absorbing phase in which the
epidemic dynamics stops.  The range of $p$ with $\rho=0$ widens for
higher contrast $w$.  Only for $p > p_{R}$ that a resulting state
of $\rho \neq 0$ with finite infection re-emerges.  The values of
$p_{L}$ and $p_{R}$ are also dependent on the network parameters
$k$ and $p$ and the SIS epidemic parameters $\lambda$ and
$\gamma$.

\begin{figure}
\begin{center}
\epsfig{figure=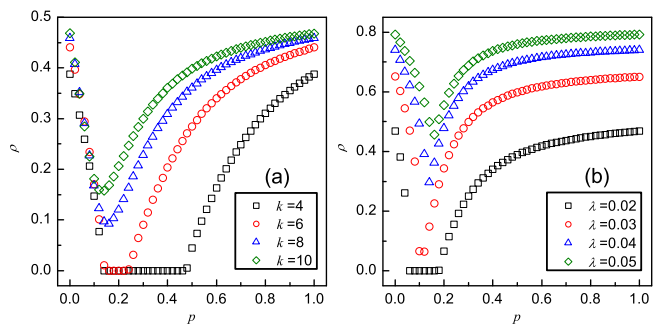,width=1.0\linewidth} \caption{The fraction
of infected nodes $\rho$ as a function of the fraction $p$ of
heavier weighted links as obtained by numerical simulations.  (a)
Results in random regular networks of different degrees $k=10$,
$8$, $6$, and $4$.  The SIS model parameters are $\lambda = 0.02$
and $\gamma=0.01$. The ratio of the weights in the links is
$w_{1}/w_{0} = 50$.  (b) Results for different values of the
infection probability $\lambda = 0.02$, $0.03$, $0.04$ and $0.05$
in random regular networks of $k=10$.  The ratio of the weights in
the links is $w_{1}/w_{0} = 10,000$, and the recovery probability
is $\gamma = 0.01$.}
    \end{center}
\end{figure}

Figure~2 shows the dependence of $\rho$ on the other parameters of
the model.  Figure~2(a) shows the behavior of $\rho(p)$ for
different values of the degree $k$ in the network structure, for
fixed $w_{1}/w_{0} = 50$, $\lambda = 0.02$ and $\gamma=0.01$.  For
small $p$, $\rho(p)$ shows only a weak dependence on $k$ before
reaching its minimum.  A larger degree suppresses the re-entrance
behavior.  As the degree of the network drops, the minimum in
$\rho(p)$ goes deeper and the value of $p$ at which the minimum
occurs shifts higher.  As the degree becomes sufficiently small,
the drop in $\rho$ at small $p$ is more pronounced so that
$\rho=0$ for $p>p_{L}$ for $k=6$ and $k=4$.  The range of $p$
between $p_{R}$ and $p_{L}$ increases as $k$ decreases.  For
larger values of $p$, a network of a higher degree gives a higher
infected fraction. Figure~2(b) shows the behavior of $\rho(p)$ for
different values of the infection probability $\lambda$ in
networks with $k=10$, $w_{1}/w_{0} = 10,000$ and $\gamma=0.01$.
The infected fraction is generally higher for a higher infection
probability. For cases in which there exists a range of $p$ with
$\rho = 0$ for a smaller value of $\lambda$ (see for example
$\lambda = 0.02$), an increased $\lambda$ will lift $\rho(p)$ up
to the extent that the $\rho=0$ state does not exist anymore.

\section{Qualitative treatment, Phase Diagram and Physical Picture}

A mean field treatment that gives qualitative agreement with
simulation results and captures all the key features can be
formulated readily.  We ignore any spatial correlation generated
by the infection process and assume that the infected nodes at a
moment in time are randomly distributed on the random regular
network.  Together with the fact that the weighted links are
randomly assigned, the probability that an infected node chooses a
susceptible neighbor for a possibility infection via a link of
weight $w_{0}$ is given according to Eq.~(1) by
\begin{equation}
{\cal P}_{w_0} = \frac{1}{{1 + (k - 1)(1 - p) + (k - 1)pw}},
\end{equation}
where $w$ is the ratio of the two weights. Similarly, the
probability that an infected node chooses a susceptible neighbor
for a possibility infection via a link of weight $w_{1}$ is
\begin{equation}
{\cal P}_{w_1} = \frac{w}{{(k - 1)(1 - p) + w + (k - 1)pw}}.
\end{equation}
A dynamical equation for the density of infected nodes $\rho$ can
then be written down as
\begin{equation}
\frac{{d\rho }}{{dt}} = (1 - \rho )\lambda  \cdot k\rho \left[ {(1
- p){\cal P}_{w_0} + p{\cal P}_{w_1}} \right] - \gamma \rho \, ,
\label{Eq:simplerho}
\end{equation}
where the first term on the right-hand-side of
Eq.~(\ref{Eq:simplerho}) describes the changes in $\rho$ due to
the infection and the second term accounts for recovery.

Solving $d\rho /dt=0$ for the density of infected nodes in the
steady state $\rho^{(MF)}$ within the mean field theory gives
\begin{equation}
\rho^{(MF)}  = 1 - \frac{1}{{k\left( {(1 - p){\cal P}_{w_{0}}  +
p{\cal P}_{w_{1}} } \right)}} \, \frac{\gamma }{\lambda} \,.
\label{Eq:MF}
\end{equation}
For the cases of $p=0$ and $p=1$, ${\cal P}_{w_{0}} = {\cal
P}_{w_{1}} = 1/k$ and thus $\rho^{(MF)}(p=0) = \rho^{(MF)}(p=1)$
as required. However, the value $\rho^{(MF)}(p=0) =
\rho^{(MF)}(p=1) = 1 - \gamma/\lambda$, which is the result of a
random regular network, is slightly higher than the simulation
results and it does not show a $k$-dependence as observed
numerically.  Despite the agreement is only qualitative, the
solution captures all the key features as we now discuss.  The
solution in Eq.~(\ref{Eq:MF}) has a factor in front of
$\gamma/\lambda$ that provides the dependence on $k$ and
$w=w_{1}/w_{0}$ for $0 < p < 1$.  For given $k$ and $w$,
Eq.~(\ref{Eq:MF}) gives
\begin{equation}
\rho^{(MF)} \approx 1 - \frac{\gamma}{\lambda} \left(1 +
\frac{(k-1)(w-1)^{2}}{k(k-1+w)} \, p \right) \label{Eq:MFlowp}
\end{equation}
in the limit of $p \rightarrow 0$.  This shows explicitly a drop
in $\rho_{s}$ from its value of $1 - \gamma/\lambda$ at $p=0$ as
$p$ increases.  Similarly, Eq.~(\ref{Eq:MF}) gives
\begin{equation}
\rho^{(MF)} \approx 1 - \frac{\gamma}{\lambda} \left(1 +
\frac{(k-1)(w-1)^{2}}{kw(1+ (k-1)w)} \, q \right)
\label{Eq:MFhighp}
\end{equation}
with $q = 1-p \rightarrow 0$ in the $p \rightarrow 1$ limit and
predicts an increase towards the value of $\rho$ at $p=1$ as $p
\rightarrow 1$.  Both features are in agreement with simulation
data.

The theory also predicts the existence of a minimum of $\rho$, as
well as a range of $p$ in which $\rho=0$ under suitable conditions
and thus the re-entrance behavior.  As ${\cal P}_{w_{0}}$ and
${\cal P}_{w_{1}}$ also depend on $p$, setting $\rho^{(MF)} = 0$
in Eq.~(\ref{Eq:MF}) gives a quadratic equation for $p$ with the
solutions
\begin{equation}
p_L  = \frac{{\gamma  - \frac{{2k(\lambda  - \gamma )}}{{w - 1}} -
\sqrt {\gamma ^2  - \frac{{4k^2 w\lambda (\lambda  - \gamma
)}}{{(k - 1)(w - 1)^2 }}} }}{{2\left( {k\lambda  - (k - 1)\gamma }
\right)}},\label{Eq:pL}
\end{equation}
\begin{equation}
p_R  = \frac{{\gamma  - \frac{{2k(\lambda  - \gamma )}}{{w - 1}} +
\sqrt {\gamma ^2  - \frac{{4k^2 w\lambda (\lambda  - \gamma
)}}{{(k - 1)(w - 1)^2 }}} }}{{2\left( {k\lambda  - (k - 1)\gamma }
\right)}}. \label{Eq:pR}
\end{equation}
The values $p_{L}$ and $p_{R}$, which depend on the network
structural parameters $k$ and $w$ as well as the SIS parameters
$\lambda$ and $\gamma$, may take on complex values.  In this case,
the theory predicts that $\rho > 0$ over {\em all} values of $p$
with a minimum at some value of $p$.  The system thus remains in an
active phase, i.e., the epidemic dynamics never stops.  An alternative
interpretation of the result is that for the whole range of $p$,
the given infection probability $\lambda$ is above the infection
threshold determined by the network structural parameters and
recovery probability and thus leading to $\rho \neq 0$.  Under
suitable conditions, $p_{L}$ and $p_{R}$ are real and the two
values separate three different regimes.  For $p<p_{L}$ and
$p>p_{R}$, the system reaches an active and mixed phase with coexisting
susceptible and infected nodes.  For $p_L\le p \le p_R$, the
system evolves into an absorbing and healthy phase ($\rho =0$) with all
the nodes being susceptible nodes, i.e., an AllS phase.  Alternatively,
the infection probability is below the corresponding threshold.  All
these features are in agreement with simulation data. As discussed
in Fig.~1 and Fig.~2, the healthy phase and the associated
re-entrance behavior emerge when $w$ is sufficiently large, for
given $k$, $\gamma$ and $\lambda$.  Using Eq.~(\ref{Eq:pL}) or
Eq.~(\ref{Eq:pR}), $p_{L}$ and $p_{R}$ will take on real values
for $w > w_{c}$, where the critical value $w_{c}$ is given by
\begin{equation}
w_c  = 1+\frac{2\left({\left(
{k^2 \lambda ^2  + \Phi}
\right) - k^2 \lambda \gamma}\right)}{{\gamma ^2 (k - 1)}},
\end{equation}
where $\Phi=\sqrt {k^2 (\gamma  - \lambda )\lambda \left(
{k^2 \lambda \gamma (1 - \lambda ) - \gamma ^2 (k - 1)} \right)}$.

\begin{figure}
\begin{center}
\epsfig{figure=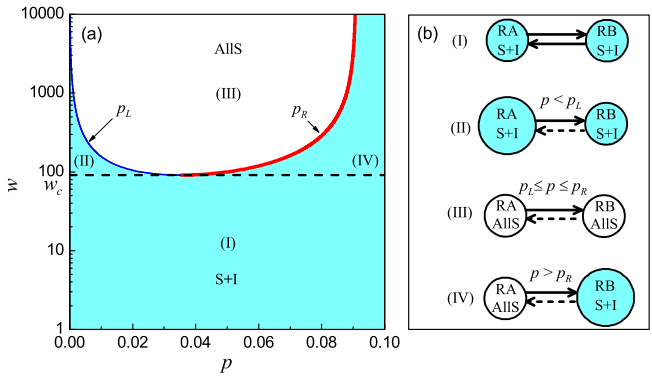,width=1.0\linewidth}
\caption{(color online) (a) Phase diagram plotted in $p$-$w$ space
for $k=10$, $\lambda=0.02$ and $\gamma=0.01$, showing four regions
(I), (II), (III), (IV) of different qualitative behavior.  The
phase boundaries (thinner blue line and thicker red line) that
separates the healthy (AllS) phase and a mixed epidemic (S+I)
phase (colored) are plotted using Eqs. (\ref{Eq:pL}) and
(\ref{Eq:pR}). A threshold $w_{c}$ (dashed line) marks two
different behavior. For $w < w_{c}$, the system is in a mixed
phase for all values of $p$ (region (I)). For $w>w_{c}$, the
system shows a re-entrance behavior as a function of $p$, starting
from a mixed phase (region (II)) at small $p$ through a healthy
phase (region (III)) for $p_{L}<p<p_{R}$ to a mixed phase (region
(IV)) again for $p>p_{R}$. (b) Schematic diagrams showing the
different mechanisms in spreading in the four regions in the phase
diagram, as explained in the text.}\label{Fig:simpletheory}
    \end{center}
\end{figure}
With a theory capable of exhibiting all the key features, it will
be a convenient tool for exploring the existence of different
phases in the parameter space.  As an example,
Fig.~\ref{Fig:simpletheory}(a) shows the possible phases in the
$p$-$w$ parameter space related to the network structure, for
fixed values of $k=10$, $\lambda=0.02$ and $\gamma=0.01$.  In this
case, $w_{c} \approx 90.88$ as indicated by the horizontal dashed
line.  For $w<w_{c}$ (see Region (I)), the system is in an active and
mixed phase labelled by S+I over the whole range of $p$.  This is the
region in which $\rho$ remains finite for all values of $p$ and
shows a minimum.  For $w >w_{c}$, there are two phase boundaries
as given by Eq.~(\ref{Eq:pL}) (see thinner line on the left) and
Eq.~(\ref{Eq:pR}) (see thicker line on the right). In between
these phase boundaries is a region (see Region (III)) in which the
system evolves into a phase with only susceptible nodes (labelled
AllS).  On either side of the AllS region are regions
corresponding to the active and mixed phase (labelled S+I) for small $p$ (see Region (II)) and large $p$ (see Region (IV)).

\begin{figure}
\begin{center}
\epsfig{figure=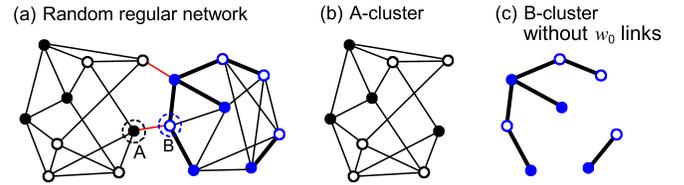,width=1.0\linewidth} \caption{(color
online) (a) A sketch of a random regular network. The open
(closed) circles represent nodes of state S (state I).  Links of
weight $w_0$ ($w_1$) are shown by the thinner (thicker) lines.
Examples of an A-type nodes (marked A) carrying only $w_{0}$ links
and a B-type node carrying has at least one $w_{1}$ link (marked
B) are shown.  Bridge-links that connect A-type and B-type nodes a
are of weight $w_{0}$ (thin red lines). (b) The A-cluster
consisting of A-type nodes and $w_{0}$ links. (c) The B-cluster
consisting of B-type nodes, but shown here with the $w_{0}$ links
removed to emphasize the importance of the weighted links when $w
= w_{1}/w_{0} \gg 1$.}\label{Fig:sketch}
    \end{center}
\end{figure}

A physical picture on the effects of an inhomogeneous link
weighting emerges from detailed analysis of simulation data and
the mean field approach.  It will be convenient for later
discussions to regard the network as consisting of three
components: two types of clusters connected by bridge links, as
shown in Fig.~\ref{Fig:sketch}(a).  Here, the thin (thick) lines
are links of weight $w_{0}$ ($w_{1}$), and the open and closed
circles are $S$ and $I$ nodes, respectively.  Nodes with {\em all}
the links having the smaller weight $w_{0}$ are called A-type
nodes.  Nodes with {\em one or more} links of the higher weight
$w_{1}$ are called B-type nodes.  An A-cluster is one that
consists of A-type nodes linked together via links of weight
$w_{0}$, as shown in Fig.~\ref{Fig:sketch}(b).  Similarly, a
B-cluster is one that consists of B-type nodes connected together.
In the case of $w=w_{1}/w_{0} \gg 1$, the weighted links $w_{1}$
play an important role as the disease would be trapped among the
weighted links once it gets into a B-cluster.
Figure~\ref{Fig:sketch}(c), therefore, shows a B-cluster with the
$w_{0}$ links removed.  By definition, the links that connect a
A-cluster and a B-cluster must be of weight $w_{0}$.  These links
are important in understanding the physics in the model and they
are called bridge-links.  The part of the network shown in
Fig.~\ref{Fig:sketch}(a) consists of the A-cluster in
Fig.~\ref{Fig:sketch}(b) and the B-cluster in
Fig.~\ref{Fig:sketch}(c) connected by two bridge-links.

This picture of the network facilitates a qualitative
understanding of the different regions in the phase diagram (see
Fig.~\ref{Fig:simpletheory}(a)).  According to Eq.~(1), once the
disease reaches the nodes in a B-cluster, the weighted links
$w_{1}$ will be preferentially selected for spreading the disease.
Although the B-clusters do not absorb disease into them, as the
bridge-links are of weight $w_{0}$, they tend to retain the
disease and this effect is increasingly important as $w_{1}/w_{0}$
increases.  Consider Regions~(II), (III), (IV) for $w
> w_{c}$ for $w_{1}/w_{0} \gg 1$.  In Region~(II) with $p < p_{L}$,
there is a dilute fraction of heavily weighted $w_{1}$ links,
forming B-clusters of small sizes.  It should be noted that the
SIS model in a sparse cluster either small in size or having a
small degree, would evolve into a state with all susceptible
(AllS) nodes, as readily seen in the extreme case of a two-node
cluster in which the disease ends when one or two infected nodes
recover.  When a few of the $w_{1} \gg w_{0}$ links are
introduced, the network structure can be viewed as a big A-cluster
in which isolated B-clusters of small sizes are embedded.  The
disease is sustained in the A-cluster, while once the disease gets
into the B-cluster, the B-type nodes will not infect the
neighboring A-type nodes again. If the B-clusters are isolated
from the background A-cluster, the disease would die out within
the B-cluster. However, the bridge-links continues to infect the
B-type nodes that are connected to the A-cluster. Therefore, the
B-cluster serves as a sink to the disease, as the disease could
only get in but not out. This process is schematically shown in
Fig.~\ref{Fig:simpletheory}(b) for Region (II), with the thick and
dashed arrows showing the asymmetry that A-type nodes would infect
B-type nodes more readily than the other way round, while the big
A-cluster is in a mixed phase and the B-type nodes at the
bridge-links have the chance of being repeatedly infected making
the small B-clusters also in a mixed phase (filled circles).
Effectively, the presence of the B-clusters serve to reduce the
tendency of spreading the disease by the A-type nodes.
Mathematically, it is represented by an effective infection
probability that is reduced from $\lambda$ by a factor given by
the terms in the parentheses in Eq.~(\ref{Eq:MFlowp}), leading to
a drop in $\rho$.  In Region (III) with $p_{L} \le p \le p_{R}$,
the B-clusters grow in size but they are not big enough to support
an epidemic within them.  They continue to be sinks for the
disease.  However, the higher fraction of $w_{1}$ links are
sufficient to reduce the size and the
average degree of the A-clusters to the extent that
the disease can no longer sustain.  Without the continual
infection via the bridge-links, the A-clusters and B-clusters
eventually reach an AllS phase.  This is schematically shown in
Fig.~\ref{Fig:simpletheory}(b) (Region (III)) with the open
circles representing the AllS phase in the A-clusters and
B-clusters.  The arrows again represent the asymmetry in infection
between A-clusters and B-clusters.  In Region (IV) with $p >
p_{L}$, the B-clusters are sufficiently large to sustain the
epidemic within them.  The effects that the B-clusters are sinks
and the A-clusters get smaller as $p$ increases put the A-clusters
in an AllS phase.  As $p$ increases, the number of healthy A-nodes
drops and $\rho$ increases.  The epidemic thus proceeds and
sustains only within the B-clusters.  The A-clusters disappear
well before the $p \rightarrow 1$ limit because a dilute fraction
of $w_{0}$ links is insufficient to form A-clusters.  As $p
\rightarrow 1$, the network structure is that of a big B-cluster
in which there are some isolated links of weight $w_{0}$.  These
isolated links have the effect of reducing the number of neighbors
that an infected B-node could choose to infect.  This effect is
represented mathematically by an effective infection probability
that is reduced from $\lambda$ by a factor given by the terms in
the parenthesis in Eq.~(\ref{Eq:MFhighp}). This is depicted in
Fig.~\ref{Fig:simpletheory}(b) (Region (IV)). There is also an
asymmetry in the formation of A-clusters and B-clusters.  While
one $w_{1}$ weighted link in a background of $w_{0}$ links leads
to the formation of a B-cluster, isolated $w_{0}$ links in a
background of $w_{1}$ links do not lead to A-nodes let alone
A-clusters.  As a result, $\rho(p)$ is not symmetrical about
$p=1/2$.

For $w<w_{c}$, the physics is basically the same.  The difference
is that the smaller contrast $w_{1}/w_{0}$ makes the asymmetry
effect in infections between A-clusters and B-clusters less
apparent.  The inhomogeneous network structure can be considered
to be effectively homogeneous with an effective infection
probability given by the denominator in the second term in
Eq.~(\ref{Eq:MF}).  The smaller contrast makes the effect of the
small clusters less important, leading to a mixed phase with
finite and non-monotonic $\rho$ over the whole range of $p$.  This
is schematically shown in Fig.~\ref{Fig:simpletheory}(b) (Region
(I)) with the arrows indicating nearly the almost symmetric mutual
infections between A-clusters and B-clusters when the contrast is
small.

\section{Quantitative treatment}

Despite the success of the mean field approach, it only gives a
qualitative understanding as much spatial correlation is ignored.
Such correlation is important because the chance of infecting a
susceptible node depends sensitively on its neighboring nodes and
their local environment.  A better theory must, therefore, take
into account of the local environment of a node.  Here, we aim at
sketching the key ideas in formulation the theory.  For a
susceptible node, we label its neighborhood by $S(k,n',n'_I,m'_I)$
when the node is connected to $n'$ neighbors through links of
weight $w_{0}$ among them $n'_{I}$ are infected {\em and} $k-n'$
neighbors through links of weight $w_{1}$ among them $m'_{I}$ are
infected. Similarly, the neighborhood of an infected node can be
labelled by $I(k,n',n'_I,m'_I)$, with the labels in the
parentheses taking on the same meaning.

\begin{figure}
\begin{center}
\epsfig{figure=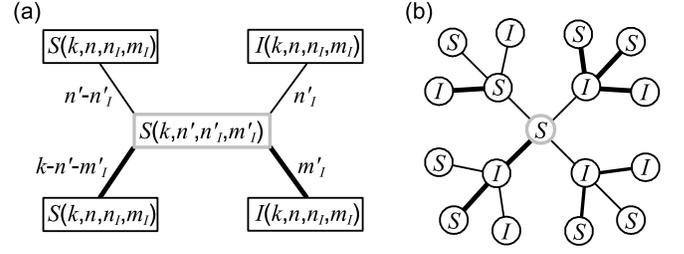,width=1.0\linewidth} \caption{A better
theory is constructed with variables being the local environment
of a node.  A susceptible node labelled $S(k,n',n_{I}',m_{I}')$
has $n'$ neighbors connected through links of weight $w_{0}$ among
them $n_{I}'$ are infected and $k-n'$ neighbors connected through
links of weight $w_{1}$ among them $m_{I}'$ are infected. (a)
Schematic diagram showing a susceptible node
$S(k,n',n_{I}',m_{I}')$ connected to neighbors of state $S$ and
$I$ and the links could be of weight $w_{0}$ (thin lines) or
$w_{1}$ (thick lines).  The neighbor's neighborhood can be
labelled accordingly.  The probability that the susceptible node
is infected in given by $\bar{\lambda}$ in Eq.~(\ref{barlambda}).
(b) An example for $k=4$ in which the susceptible node in the
middle has an environment $S(4,3,2,1)$ and its four neighbors have
their neighborhood described by $I(4,1,0,1)$, $S(4,3,1,1)$,
$I(4,2,1,0)$, and $I(4,2,0,1)$, respectively. }\label{Fig:X1}
    \end{center}
\end{figure}

The probability that a susceptible node will be infected depends
not only on its neighborhood but also on its infected neighbors'
local environment.  Figure~\ref{Fig:X1}(a) illustrates the
different possible combinations that an $S$ node would encounter.
An example is given in Fig.~\ref{Fig:X1}(b) for $k=4$, where the
$S$ node in the middle has a local environment described by
$S(4,3,2,1)$, and its four neighbors are described by $I(4,1,0,1)$
(upper right), $S(4,3,1,1)$ (upper left), $I(4,2,1,0)$ (lower
left), and $I(4,2,0,1)$ (lower right).  As the system evolves, the
numbers of nodes described by $S(k,n';n'_I,m'_I)$ and
$I(k,n';n'_I,m'_I)$ also evolve.  These numbers form the variables
of the theory.  To close the dynamical equations for these
variables, we assume a random distribution of neighbors around a
node as follows.

Considering an $S$ node of neighborhood $S(k,n',n'_I,m'_I)$, the
probability that it is connected to an infected neighbor of
neighborhood $I(k,n,n_I,m_I)$ through a link of weight $w_{0}$ is
assumed to be
\begin{equation}
q_{I,w_0 }(k,n,n_I ,m_I ) = \frac{(n - n_I )f_I (k,n,n_I ,m_I
)}{\Omega_{I,w_0 }}. \label{Eq:qIw0}
\end{equation}
Here, $f_I (k,n,n_I ,m_I )$ is the fraction of nodes in the local
environment $I(k,n,n_I ,m_I )$ in the system and $\Omega_{I,w_0 }
= \sum\nolimits_{n,n_I ,m_I } {(n - n_I )f_I (k,n,n_I ,m_I )}$ is
a normalization factor.  Similarly, the probability that it is
connected to an infected neighbor of neighborhood $I (k,n,n_I ,m_I
)$ through a link of weight $w_1$ is given by
\begin{equation}
q_{I,w_1}(k,n,n_I ,m_I ) = \frac{(k - n - m_I )f_I (k,n,n_I ,m_I
)}{\Omega_{I,w_1}}, \label{Eq:qIw1}
\end{equation}
where the normalization factor $\Omega_{I,w_1 } =
\sum\nolimits_{n,n_I ,m_I } {(k - n - m_I )f_I (k,n,n_I ,m_I )}$.
Following a similar consideration, the probability that the $S$
node is connected to a susceptible neighbor of neighborhood
$S(k,n,n_I ,m_I )$ through a link of weight $w_0$ is given by
$q_{S,w_0 }(k,n,n_I,m_I) = (n - n_I )f_S (k,n,n_I ,m_I
)/\Omega_{S,w_0 }$.  Here, $f_S (k,n,n_I ,m_I)$ is the fraction of
nodes in the local environment $S(k,n,n_I ,m_I )$ in the system
and $\Omega _{S,w_0 }=\sum\nolimits_{n,n_I ,m_I } {(n - n_I )f_S
(k,n,n_I ,m_I )}$.  Similarly, we have $q_{S,w_1} (k,n,n_I ,m_I )
= (k - n - m_I )f_S (k,n,n_I ,m_I )/\Omega_{S,w_1 }$ for
connecting to a susceptible neighbor of environment $S(k,n,n_I
,m_I )$ through a link of weight $w_{1}$, with $\Omega _{S,w_1 }
= \sum\nolimits_{n,n_I ,m_I } {(k - n - m_I )f_S (k,n,n_I ,m_I
)}$.

According to Eq.~(1), an infected neighbor of neighborhood
$I(k,n,n_I ,m_I )$ has a probability
\begin{equation}
{\cal P}_{w_{0}}(k,n) = \frac{w_0}{nw_0 + (k - n)w_1}
\label{Eq:Pw0}
\end{equation}
to choose the $S$ node for infection if the link is of weight
$w_{0}$, and a probability
\begin{equation}
{\cal P}_{w_{1}}(k,n) = \frac{w_1}{nw_0  + (k - n)w_1}
\label{Eq:Pw1}
\end{equation}
if the link is of weight $w_{1}$. Taking into account of all
possible cases of infected neighbors linked through weight $w_{0}$
links, the average probability $\overline{\cal P}_{w_0}$ that the
$S$ node is selected for infection through a link of weight
$w_{0}$ is
\begin{equation}
\overline{\cal P}_{w_0 } (k) =  {\sum\limits_n {\sum\limits_{n_I }
{\sum\limits_{m_I } {q_{I,w_0 } (k,n,n_I ,m_I ){\cal P}_{w_0 }
(k,n)} } } }.
\end{equation}
Similarly, the average probability $\overline{\cal P}_{w_1}$ the
the $S$ node is selected for infection through a link of weight
$w_1$ is
\begin{equation}
\overline{\cal P}_{w_1 }(k)  = {\sum\limits_n {\sum\limits_{n_I }
{\sum\limits_{m_I } {q_{I,w_1 } (k,n,n_I ,m_I ){\cal P}_{w_1 }
(k,n)} } } }.
\end{equation}
Finally, the probability of the $S$ node is infected is given by
\begin{equation}
\bar \lambda (k,n',n'_I ,m'_I ) = 1 - (1 - \lambda \overline{\cal
P}_{w_0 }(k) )^{n'_I } (1 - \lambda \overline{\cal P}_{w_1 }(k)
)^{m'_I },\label{barlambda}
\end{equation}
where the second term gives the probability that the node is not
infected by any one of the $(n_{I}'+m_{I}')$ infected neighbors
via links of either weight $w_{0}$ or $w_{1}$.

The dynamical equations of the variables $f_S(k,n,n_I,m_I)$ can
now be written down as
\begin{eqnarray}
\frac{{df_S (k,n,n_I ,m_I )}}{{dt}}= &-& f_S (k,n,n_I ,m_I )\bar
\lambda (k,n,n_I ,m_I )\nonumber \\
&+& f_I (k,n,n_I ,m_I )\gamma\nonumber \\
&+& \delta f_S
(k,n,n_I ,m_I ).\label{dynamiceqofS}
\end{eqnarray}
The first term in the right-hand-side accounts for the drop in the
fraction $f_S (k,n,n_I ,m_I )$ when susceptible nodes with
neighborhood $S(k,n,n_I ,m_I )$ are infected.  The second term
accounts for the recovery of infected nodes of neighborhood
$I(k,n,n_I ,m_I )$.  When an $S$ node is infected, its neighboring
susceptible nodes have a modified neighborhood with one more
infected neighbor and the third term accounts for such a change in
$f_S (k,n,n_I ,m_I )$.  As an example, consider the infection of
the $S$ node in the middle of Fig.~\ref{Fig:X1}(b). In this case,
the first term in Eq.~(\ref{dynamiceqofS}) accounts for the drop
of $1/N$ in the fraction $f_S(4,3,2,1)$ when infection occurs with
the probability $\bar\lambda$.  At the same time, the upper left S
node has its neighborhood changed from $S(4,3,1,1)$ to
$S(4,3,2,1)$, leading to an increase of $1/N$ given by the third
term in Eq.~(\ref{dynamiceqofS}) when it is applied to
$f_S(4,3,2,1)$.

The dynamical equations of the variables $f_I(k,n,n_I,m_I)$ can
also be written down as
\begin{eqnarray}
\frac{{df_I (k,n,n_I ,m_I )}}{{dt}}=&+&f_S (k,n,n_I ,m_I )\bar
\lambda (k,n,n_I ,m_I )\nonumber \\
&-& f_I (k,n,n_I ,m_I )\gamma\nonumber \\
&+& \delta f_I(k,n,n_I ,m_I ),\label{dynamiceqofI}
\end{eqnarray}
with the terms carrying similar meanings as in
Eq.~(\ref{dynamiceqofS}).  Equations~(\ref{dynamiceqofS}) and
(\ref{dynamiceqofI}) form a closed set of equations for the
variables $f_S(k,n,n_I,m_I)$ and $f_I(k,n,n_I,m_I)$.  The theory
accounts for spatial correlation up to the neighborhoods of the
nearest neighbors of a node.  In contrast, the simple mean field
approach is a site-approximation that ignores any spatial
correlation.  Incorporating a longer spatial correlation is often
necessary in problems in which the evolution is related to
comparing the states of neighboring nodes, as in Refs.
\cite{Noel,Ji,Demirel}. The set of equations can be iterated in
time to study the evolution of the systems.  Steady state
properties can be studied by either setting the equations to zero
or iterating the equations to the long time limit.  The fraction
of infected nodes in the steady state can be found by
\begin{equation}
\rho = \sum\limits_n {\sum\limits_{n_I } {\sum\limits_{m_I } {f_I
(k,n,n_I ,m_I )} } },
\end{equation}
which combines the steady state values of $f_I(k,n,n_I,m_I)$ for a
random regular network of degree $k$.

\begin{figure}
\begin{center}
\epsfig{figure=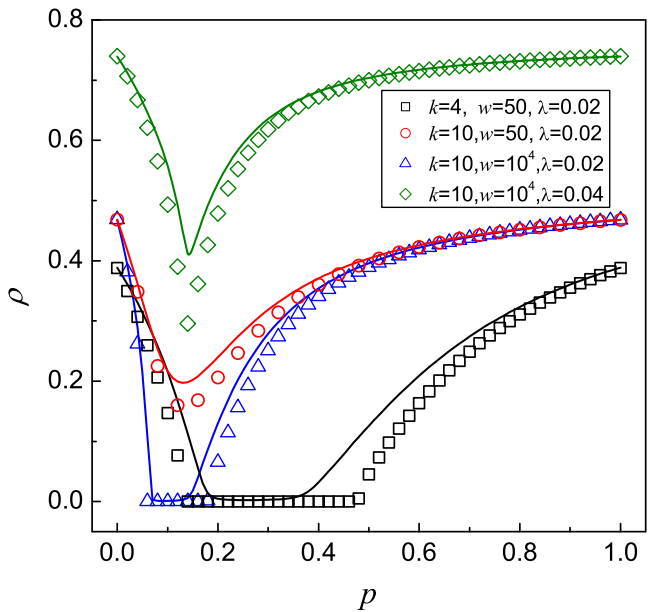,width=1.0\linewidth} \caption{(color
online) The fraction of infected nodes $\rho$ as a function of the
fraction $p$ of heavier weighted links as calculated by theory
(curves) based on the dynamics of local environment of nodes.
Results for different values of $k$, $w$ and $\lambda$ are shown.
Simulation results (symbols) are included for comparison.}
\label{Fig:X2}
    \end{center}
\end{figure}

Applying Eqs.~(\ref{dynamiceqofS}) and (\ref{dynamiceqofI}) to our
model and solving them numerically for different degrees $k$ and
infection probabilities gives the results as shown in
Fig.~\ref{Fig:X2}.  The theory agrees reasonable well with
simulation data.  In particular, the theory is in quantitative
agreement with simulation data for both $p \ll 1$ and for a large
part of $p$ in which $\rho$ increases.  For cases in which $\rho$
goes through a minimum without vanishing, the theory captures the
behavior very well, except missing the depth of the minimum.
Discrepancies remain, however, in the vicinity where $\rho$
vanishing continuously near $p_{L}$ and $p_{R}$.  The theory
predicts a range of $p_{L} \le p \le p_{R}$ with $\rho=0$, but the
values of $p_{L}$ and $p_{R}$ are slightly off.  The discrepancy
shows that the effects of fluctuations in the local environments
of S nodes and I nodes near the absorbing transitions are so
important that even spatial correlation to the extent that our
theory includes is insufficient to capture the behavior near the
transitions accurately.  Although one could, in principle, include
longer spatial correlation by using a bigger set of variables, the
theory will involve more complicated and many more equations. The
inclusion of spatial correlation up to the neighborhood of nearest
neighbors as in the present theory represents a good balance
between accuracy and complexity of the theory.

\section{Summary}

We proposed and studied in detail the SIS epidemic model with a contact
infection process in a
network consisting of two different kinds of links mimicking two
kinds of relationship.  For an infected nodes with $k$ links, a
link of heavier weight $w_{1}$ has a higher chance of being chosen
as the path of infection than a link of weight $w_{0}$.  Detailed
numerical simulations revealed that the fraction of infected nodes
$\rho(p)$ varies with the fraction of $w_{1}$ weighted links $p$
non-monotonically.  For small contrasts of $w = w_{1}/w_{0}$,
$\rho(p)$ shows a minimum at a particular value of $p$ that is
dependent on the degree $k$ and SIS model parameters.  This
signifies an optimal suppression of the epidemic.  For
sufficiently large contrasts, the suppression leads to a range
$p_{L} \le p \le p_{R}$ in which the disease will eventually die
off and give rise to a 100\% healthy state.  This leads to a
re-entrance behavior as a function of $p$, with the system goes
through an active epidemic phase, an absorbing healthy phase and an active epidemic phase again as $p$ increases.  We also studied the
effects of different values of $k$ and the infection probability.

Attempts to explain the numerical features proved to be
non-trivial.  A mean field theory that ignores spatial correlation
was formulated.  The theory captures all the key features in
simulation results, but the agreement is only qualitative.  It
gives explicit analytic expressions for $\rho(p)$ in the $p
\rightarrow 0$ and $p \rightarrow 1$ limits.  The limiting values
show that the infection probability is effectively reduced in the
two limiting cases of having a dilute fraction of $w_{1}$ links in
a background of $w_{0}$ links and vice versa, but the reduction is
asymmetric.  In addition, it also exhibits the re-entrance
behavior when $w > w_{c}$ and gives an explicit expression for
$w_{c}$.  A merit of the theory is that it leads to a transparent
physical picture.  The picture emphasizes the importance of
clusters formed by nodes with links with the higher weight.  These
links confine the spread to within a cluster.  At small $p$, the
clusters are small and the disease cannot sustain in them.  The
disease remains only among the nodes with only $w_{0}$ links and
the infection probability is effectively reduced.  At large $p$,
there is a big cluster and infection can be sustained in it.
Replacing some $w_{1}$ links by $w_{0}$ links reduces the choice
of neighbors of the infected nodes and thus effectively reduces
the infection probability.  For sufficiently large $w_{1}/w_{0}$,
a state of all healthy nodes can be reached for appropriate values
of $p$ when the cluster sizes are big enough to disrupt the
epidemics via $w_{0}$ links and yet they are not big enough for
sustaining the epidemics via $w_{1}$ in the clusters.

An improved theory that uses the local environment of a node as
variables was also formulated.  Comparing to the simple mean field
theory, it has a large set of variables as a node could be
susceptible or infected and it could be connected by $k$ nodes of
in either the $S$ or the $I$ state via $w_{0}$ or $w_{1}$ links.
Amounting for the processes that would lead to a change in the
number of different local environments, dynamical equations can be
constructed for these variables.  It was shown that solving the
equations in the steady state gives results in good quantitative
agreement with simulation results.  From the epidemic dynamics,
the inclusion of a longer spatial correlation than the single-site
mean field theory is necessary.  Our work showed that including a
biased selection of infection targets in the standard SIS epidemic
model leads to a suppressed spread, even to the extent that a
healthy state may result.  The model studied here can be readily
generalized to other types of networks, as well as other types of
weight distributions among the links.  In the present model, the
infection probability $\lambda$ is taken to be independent of the
weight of the link in which the infection takes place.  An
alternative is to make use of Eq.(1) to drive both the selection
of target and infection.  Analytically, the quantitative treatment
based on the local environment provides a general framework of
formulating a reliable theory for other problems where the change
in the state of a node is related to its neighborhood and the
inclusion of a spatial correlation longer than a single site is
unavoidable.

\acknowledgments {One of us (Elvis H.W. Xu) acknowledges the
support from a Hong Kong PhD Fellowship awarded by the Research Grants Council of the Hong Kong SAR Government.  M. Tang
acknowledges the support from the National Natural Science
Foundation of China (Grant Nos.~11105025, 91324002).  Y. Do
acknowledges the support by the Basic Science Research Program
through the National Research Foundation of Korea (NRF) funded by
the Ministry of Education, Science and Technology
(NRF-2013R1A1A2010067). P.M.H. acknowledges the support of a
Direct Grant of Research from the Faculty of Science at the
Chinese University of Hong Kong in 2013-14.}

\end{document}